\documentclass{icrc2009}

\usepackage{graphicx}   
\usepackage[caption=false]{caption}    
\usepackage[font=footnotesize]{subfig} 
\usepackage{fixltx2e}
\usepackage{url}
\usepackage{textcomp}
\usepackage{amsfonts}
\usepackage{amssymb}

\newcommand{\shorttitle}[1]%
{\markboth{Proceedings of the 31\MakeLowercase{$^{st}$} ICRC, {\L}\'{o}d\'{z} 2009}{#1} }
\newcommand{\etal}{\MakeLowercase{\textit{et al. }}} 
\newcommand{\td}{\textdegree}
\newcommand{\be}{\begin{enumerate}}
\newcommand{\ee}{\end{enumerate}}

\begin{document}
\title{Observations of Supernova Remnants and Pulsar Wind Nebulae: A VERITAS Key Science Project}

\author{\IEEEauthorblockN{Brian Humensky\IEEEauthorrefmark{1} for the VERITAS Collaboration\IEEEauthorrefmark{2}}
                            \\
\IEEEauthorblockA{\IEEEauthorrefmark{1}University of Chicago Enrico Fermi Institute, Chicago, IL 60637, USA (humensky@uchicago.edu)}
\IEEEauthorblockA{\IEEEauthorrefmark{2}see R.A. Ong et al (these proceedings) or http://veritas.sao.arizona.edu/conferences/authors?icrc2009}
}

\shorttitle{T. B. Humensky \etal Observations of SNRs and PWNe with VERITAS}
\maketitle

\begin{abstract}
The study of supernova remnants and pulsar wind nebulae was one of the Key Science Projects for the first two years of VERITAS observations. VERITAS is an array of four imaging Cherenkov telescopes located at the Whipple Observatory in southern Arizona. Supernova remnants are widely considered to be the strongest candidate for the source of cosmic rays below the knee at around $10^{15}\ \textrm{eV}$. Pulsar wind nebulae are synchrotron nebulae powered by the spin-down of energetic young pulsars, and comprise one of the most populous very-high-energy gamma-ray source classes. This poster will summarize the results of this observation program. \\
  \end{abstract}

 
\section{Introduction}
The relics of supernova explosions, from the high-velocity blast waves to the rapidly spinning, highly magnetic compact stars produced in some events, play a crucial role in high-energy astrophysics. In addition to seeding the Galaxy with metal-rich stellar ejecta, the high Mach-number shocks in supernova remnants (SNRs) are almost certainly the principal source of cosmic-ray acceleration to energies approaching the knee of the cosmic-ray spectrum. However, while the supernova birthrate and overall energetics appear adequate to yield the observed energy density of cosmic rays, it remains unclear exactly how the acceleration occurs. Radio observations of SNRs demonstrate clearly that electrons have been accelerated up to $\sim\textrm{GeV}$ energies, but it is only in the past decade that X-ray observations have demonstrated the presence of electrons with energies up to $\sim 100\ \textrm{TeV}$. However, it is the ions that dominate both the energy density of the cosmic rays and the dynamical evolution of the SNR shocks.  Therefore, it is of fundamental importance that, to date, there is only ambiguous evidence of ion acceleration in SNR shocks. Detections of very-high-energy (VHE, $E>100\ \textrm{GeV}$) gamma rays from SNRs offer the opportunity for direct detection of the ion component through the decay of neutral pions that have been produced in energetic proton-proton collisions; these collisions should be particularly evident in the vicinity of dense molecular clouds \cite{Butt2003a}. By combining the TeV spectrum with that in the radio and X-ray bands, this process can be compared with inverse-Compton scattering models to determine whether the TeV emission is actually associated with the electrons or the ions. In either case, strong constraints can be placed on the acceleration process (see, e.g., \cite{Aharonian2007a}).

For supernova events that result from the collapse of massive stars, the relic neutron stars that are typically formed are sources of extremely energetic radiation as well. Particles accelerated in the pulsar magnetosphere stream outward, accompanied by Poynting flux from the rotating magnetic field, and filling a bubble whose expansion is restricted by the surrounding ejecta. As the energetic particle wind meets this restricted flow, a wind termination shock forms, at which additional acceleration occurs. Curiously, while models for particle acceleration in the magnetosphere predict a wind in which the particles carry only $\sim10^{-4}$ of the energy flux, the spectrum and dynamics of the downstream wind nebula require a particle-dominated wind. The composition of the wind changes drastically between the pulsar light cylinder and the region downstream of the termination shock, but how this happens is not yet understood. The broadband spectrum of the particles in the nebula strongly constrains the process by which this conversion occurs. For all pulsar wind nebulae (PWNe), there is a change in spectral slope between the radio and X-ray bands. Whether this change is due to a simple break due to synchrotron aging, or to a more complicated electron spectrum is not known for most PWNe. Yet, at least for those with somewhat low magnetic fields, emission in the TeV band holds crucial information. This is because the electrons that produce synchrotron radiation in the ultraviolet band, which probes the region below the steeper X-ray spectrum (but is virtually always unseen because of Galactic absorption), also produce TeV gamma rays through inverse-Compton scattering of the microwave background. Observations of PWNe at TeV energies thus offer an opportunity to investigate a key portion of the spectrum and place constraints on the acceleration and evolution of particles in these energetic systems. 

VERITAS defined four Key Science Projects for its first two years of operations, in the areas of Active Galactic Nuclei, the search for dark matter, a survey of the Cygnus region of the Galactic plane, and the study of supernova remnants and pulsar wind nebulae.  These were recognized as areas addressing high-priority science questions, questions for which specific targets or observing strategies and a substantial investment of observing time would be required.  This paper summarizes the  state of the VERITAS program of dedicated observations of SNRs and PWNe prior to the ICRC.  It complements the VERITAS survey of the Cygnus region, which contains a number of other interesting SNRs and PWNe such as $\gamma$Cygni and CTB 87.  This survey spans Galactic longitude $67 < l < 82$ and latitude $-1 < b < 4$ and is discussed elsewhere in these proceedings \cite{Weinstein2009a}.

\section{The Observing Program}
VERITAS \cite{Holder2006a} consists of four $12\textrm{-m}$ telescopes located at an altitude of $1268\ \textrm{m}$ a.s.l. at the Fred Lawrence Whipple Observatory in southern Arizona, USA (31\td\ 40' 30'' N, 110\td\ 57' 07'' W). Each telescope is equipped with a 499-pixel camera of $3.5\textrm{\td}$ field of view.  The array, completed in the spring of 2007, is sensitive to a point source of 1\% of the steady Crab Nebula flux above $300\ \textrm{GeV}$ at $5\ \sigma$ in less than 50 hours at 20\td\ zenith angle.

Candidate SNRs have been selected for observation based on one of two primary criteria:
\begin{itemize}
\item Presence of nonthermal X-ray emission (Cas A)
\item Interaction with molecular clouds, especially as indicated by maser emission (IC 443, CTB 109)
\end{itemize}
In addition, FVW 190.2+1.1 was observed as an exploration of a potential new class of VHE emitting SNRs.

PWNe have been selected primarily on the basis of their spin-down luminosity, favoring nearby energetic pulsars; these results were previously presented in \cite{Aliu2009a}.  In addition, Geminga was selected because of its proximity, strength as a gamma-ray pulsar, and detection by MILAGRO at multi-TeV energies \cite{Finnegan2009a}.

\section{Discussion}
Table~\ref{table1} summarizes the observing program to date.  The results include detections of the Crab Nebula and the SNRs Cassiopeia A and IC 443.  Point-source upper limits have been set on a number of other potential VHE sources.  The Cas A results are preliminary and reflect those shown in \cite{Humensky2009a}; final results on Cas A and additional new results \cite{Humensky2009b,Bugaev2009a,Wakely2009a} will be presented at the ICRC.  Below we discuss the individual objects.

  \begin{table*}[th]
  \caption{Summary of supernova remnant and pulsar wind nebula observations.  The measured flux or limit s indicated.  The columns in the table give the name of the object, its class, the livetime of the observations, the significance and resulting flux or limit (99\% confidence level) above a threshold of $300\ \textrm{GeV}$ (unless otherwise noted), the VERITAS reference from which the measurement is taken, and any notes about the object..  The Crab significance is quoted for 3.3 hours of 3-telescope data \cite{Celik2008a}.} 
  \label{table1}
  \centering
  \begin{tabular}{|l|c|c|c|c|c|c|}
  \hline
   Object & Class & Livetime & Significance & Flux or Limit & Ref & Notes \\
 & & (hrs) & ($\sigma$) & ($\times 10^{-12}\ \textrm{cm}^{-2}\ \textrm{s}^{-1}$) & & \\
   \hline 
Crab Nebula & PWN & 19.0 & 56 & $151 \pm 6_{stat} \pm 30_{sys}$& \cite{Celik2008a} & \\
PSR J0205+6449 & PWN & 12.8 & 1.1 & $< 2.9$ & \cite{Aliu2009a} & 3C 58 \\
PSR J0631+1036 & PWN & 13 & 0.3 & $< 1.6$ & \cite{Aliu2009a} & \\
PSR J0633+1746 & PWN & 10.4 & -0.5 & $< 2.0$ & \cite{Finnegan2009a} & Geminga \\
PSR B0656+14 & PWN & 9.4 & -1.8 & $< 0.3$ & \cite{Aliu2009a} & \\
PSR J1740+1000 & PWN & 10.5 & 0.2 & $< 1.2$ & \cite{Aliu2009a} & \\
SNR G111.7-2.1 & Shell & 20.3 & 9.8 & $4.5 \pm 0.5_{stat} \pm 0.9_{sys}$ & \cite{Humensky2009a} & Cas A (prel.) \\
SNR G189.1+3.0 & Composite & 37.9 & 8.3 & $4.63 \pm 0.90_{stat} \pm 0.93_{sys}$  & \cite{Acciari2009a} & IC 443 \\
SNR G109.1-1.0 & Shell & 4.3 & 1.5 & $F(E>400\ \textrm{GeV}) < 2.5$  & \cite{Guenette2009a} & CTB 109 \\
FVW 190.2+1.1 & FVW & 18.4 & 0.1 & $F(E>500\ \textrm{GeV}) < 0.36$ & \cite{Holder2009a} &  \\
  \hline
  \end{tabular}
  \end{table*}
 
{\bf Crab Nebula:} This is the standard candle for VHE astronomy, and VERITAS observations of the Crab Nebula reproduce historical measurements.

{\bf PSR J0205+6449:} Better known as 3C 58, this PWN has a spin-down luminosity of $2.5 \times 10^{37}\ \textrm{erg/s}$ and is among the most energetic pulsars in the galaxy.  

{\bf PSR J0631+1036:} A pulsar with a spin-down luminosity of $1.7 \times 10^{35}\ \textrm{erg/s}$ and a distance of $6.55\ \textrm{kpc}$.

{\bf PSR J0633+1746:} Also known as Geminga, this pulsar is $3\times 10^5$ years old but is only $\sim 160\ \textrm{pc}$ distant.  Its proximity may allow it to play a role in the GeV positron excess.  Very extended ($\sim3$\td) emission has been observed by MILAGRO with a median energy of $\sim 35\ \textrm{TeV}$.

{\bf PSR B0656+14:} Another nearby pulsar, located (in projection) near the center of the Monogem Ring.

{\bf PSR J1740+1000:} A pulsar with a spin-down luminosity of $2.3 \times 10^{35}\ \textrm{erg/s}$ and a distance as small as $1.36\ \textrm{kpc}$.

{\bf Cassiopeia A:} Initially detected in VHE by HEGRA \cite{Aharonian2001a} in a 232-hour observation, Cas A was confirmed as a VHE source by MAGIC \cite{Albert2007b} at the $5.3\textrm{-}\sigma$ level in 47 hours.  A final analysis of the VERITAS data set, including a measurement of the spectrum and an upper limit on the extension of the emission, will be presented at the ICRC.

\begin{figure}[!t]
  \centering
  \includegraphics[width=2.5in]{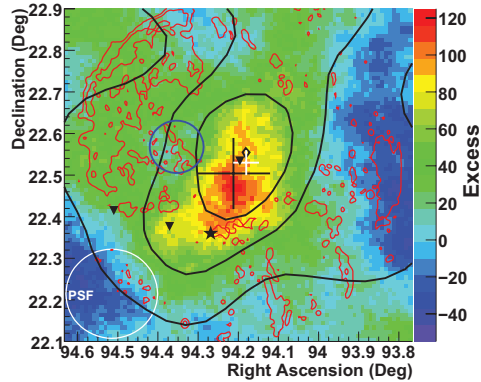}
  \caption{Inner $0.8\textrm{\td}$ of the acceptance-corrected excess map for the IC 443 field. The black cross-hair indicates the centroid position and its uncertainty (statistical and systematic added in quadrature), and the white cross-hair likewise indicates the position and uncertainty of MAGIC J0616+225 \cite{Albert2007a}.  Red contours: optical intensity \cite{McLean2000a}.  Thick black contours: CO survey \cite{Huang1986a}; black star: PWN CXOU J061705.3+222127 \cite{Olbert2001a}; open blue circle: 95\% confidence radius of 0FGL J0617.4+2234 \cite{Abdo2009a}; and filled black triangles: locations of OH maser emission (\cite{Claussen1997a}, \cite{Hewitt2006a}, J. W. Hewitt, private communication). The white circle indicates the PSF of the VERITAS array.}
  \label{fig1}
 \end{figure}

{\bf IC 443:} Detected in VHE by both MAGIC \cite{Albert2007a} and VERITAS in 2007, deep VERITAS observations have revealed that the emission is extended \cite{Acciari2009a}.  A hint that the emission is displaced from the location of the Fermi bright source 0FGL J0617.4+2234, as shown in Figure~\ref{fig1}, suggests that the morphology may be energy dependent.  This will be studied in further VERITAS observations and may be key to determining whether the VHE emission is associated with the PWN CXOU J061705.3+222127 or with the site of the SNR shock / molecular cloud interaction.

{\bf SNR G109.1-1.0:} Also known as CTB 109, this remnant is interacting with a molecular cloud on its eastern rim.  No emission was observed in a brief observation.

{\bf FVW 190.2+1.1:} Forbidden Velocity Wings may be the vestiges of very old supernova remnants.  Motivated by the possible association of HESS J1503-582 with an FVW \cite{Renaud2008a}, VERITAS observed FVW 190.2+1.1 and set a stringent upper limit at the level of 1\% of the Crab Nebula flux above $500\ \textrm{GeV}$.


\section{Summary}
VERITAS has conducted an observing program for supernova remnants and pulsar wind nebulae over the last two years, and it has born fruit with detections of Cassiopeia A and IC 443 and interesting upper limits on a number of other objects.  New results beyond those presented here will be shown at the ICRC.  With the relocation of one telescope this summer \cite{Otte2009a}, leading to an improved sensitivity and angular resolution, the future looks promising as well.

\section{Acknowledgements}
This research is supported by grants from the US Department of Energy, the US National Science Foundation, and the Smithsonian Institution, by NSERC in Canada, by Science Foundation Ireland, and by STFC in the UK. We acknowledge the excellent work of the technical support staff at the FLWO and the collaborating institutions in the construction and operation of the instrument.

\end{document}